# A Standing Stone and its Possible Astronomical Alignment

## Using Seasonal Shadow and Light Displays in the Neolithic

Dr. Daniel Brown

Sundials and gnomons pointing out dates and times are a beautiful reminder of how we understand the movement of the Sun and use it to structure day and year. They also encapsulate the basic astronomical knowledge and concepts of the individuals and therefore their societies which have built them. At the heart of every sundial lies the recognition that the daily and seasonal change of light and shadow is repeated over and over again, in the past, present and into the distant future. Such knowledge was developed well before any written documents existed during times when observing did not mean measuring but rather watching and embracing the findings into a far deeper ritualistic religious frame work. A 4,000 year old monolith located at Gardom's Edge with its striking orientation and in the midst of a landscape rich in other monuments is presented as a possible example of such astronomical knowledge and how it might have been applied to create a deeper meaning for this monument.

Within the Peak District National Park there are many ancient monuments ranging from the Neolithic, through to Roman and more modern times. Gardom's Edge close to Baslow is a special example of how in only a square kilometer you can experience the entire impact humans have had on the landscape. The gritstone scarp of Gardom's Edge consists of a medium high plateau overlooking the river Derwent. The area is littered with many rocks and overgrown with heather, bracken and partly birch trees. But it allows beautiful vistas towards the West and part of Chatsworth estate. The easiest approach in the sometimes quite boggy ground is from the free car park at the Robin Hoods Inn and outlined in Figure 1. Walking towards Baslow along the main road there is a foot path that can be taken to slowly walk higher past old braided hollow ways, cairns, and a medieval hermit's hovel. After having passed through an old dry stone wall gate the path forks and the right path along the dry stone wall will take you onto the plateau. After following the wall for a while you will cross the tumbled remains of another wall and a gate will be to the right. Entering, one is now only meters away from the monolith. Detailed descriptions of walks on site can be found in Harris[1] or Johnstone[2].

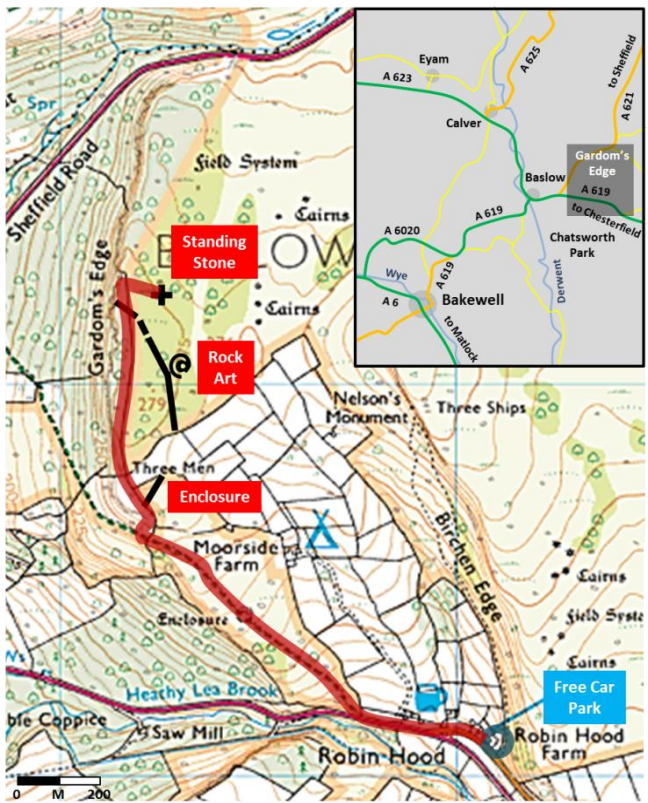

Fig 1. This OS-map illustrates the location of the standing stone at Gardom's Edge. The free car park and various relevant ancient sites are labelled, including the easiest access route in red. The inset gives a larger overview of the region.

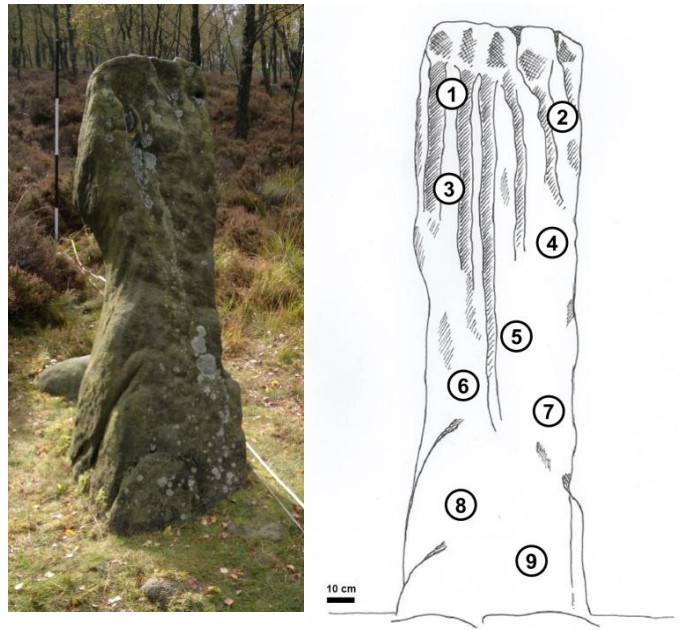

Fig. 2. A contemporary image of the standing stone (left) at midday on 13 October with the north-facing side in shadow. Note the presence of possible packing stones at the base. The sketch of the north-facing side (right) shows the erosion features as well as the nine locations used for the survey of the gradient and orientation.

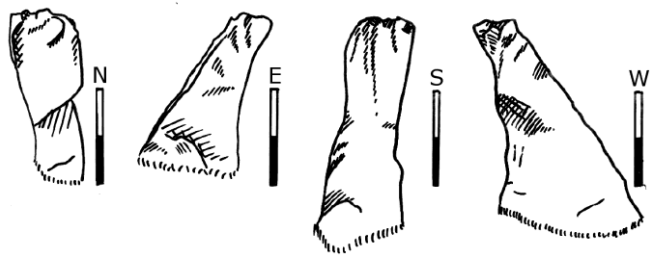

*Fig. 3. The four sketches show the profile of the monolith when looking N, E, S, and W. A meter scale is given for comparison. Note the rather unusual angular or triangular shape of the stone.*

The area surrounding the monolith contains several Neolithic monuments that are proposed to be contemporary, these include an enclosure of 100m x 500m size with walls 5-10m wide and up to 1.5m high (which will have been crossed twice while approaching the site) and in-situ rock art on an earth fasten rock 2m x 1m (located only 200m further from the monolith). More difficult to spot are some bronze age round houses and field systems. A more in depth overview of the site has been given by Barnatt[3]. These examples illustrate how humans have inhabited and farmed this region as well as given it a deeper ritual or religious meaning as expressed for example in the enclosure. Even though the entire landscape is littered with rocks the monolith is outstanding in its triangular shape and height of 2m illustrated in Figure 3. It is also located a mere 50m North-East of one of the two entrances of the enclosure. It is made from a similar gritstone as the bedrock and the surrounding boulders but it is more angular than the others. It has suffered severe weathering leading to deep localised erosion. At its base several smaller stones can be made out (see Figure 2) as well as a slight local increase in the ground level, both leading to the possible interpretation of packing stones and therefore an intentional erection of this stone. Such standing stones are rare in the Peak District as they tend to have been removed to be used in buildings over the millennia. There are no other such examples at Gardom's Edge or other adjacent moors, making this an ideal opportunity to study such a monolith in its original setting.

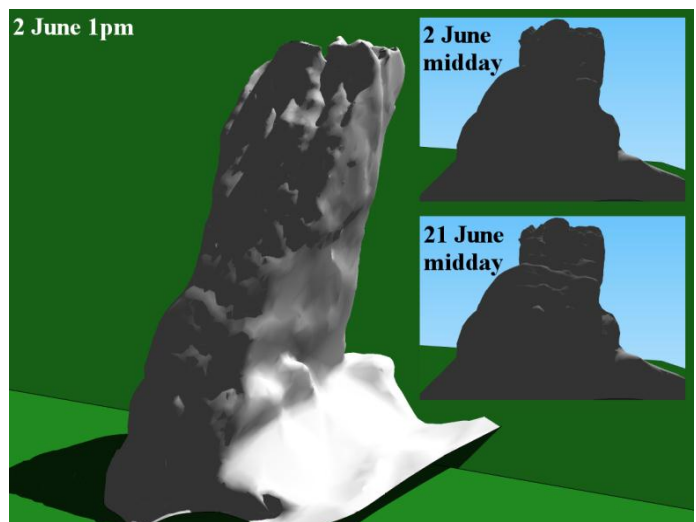

*Fig. 4. The rendered three dimensional model of the standing stone as illuminated 2,000 BC on 2 June 1pm illustrates the realistic erosion features. The top inset shows the first illumination of the north-facing side at midday and the bottom inset shows the illumination of the north-facing side at mid-summer midday. Both insets are viewing the north-facing side looking up the slope.*

A detailed erosion study was carried out for the most affected north-facing side. The systematic mapping of weathering features indicates that the stone has been at its current orientation for a significant period. The features are consistent with erosional processes at this orientation and have also occurred in other Neolithic stone circles consisting of sandstone (see Duddo stones NT931437 described by Younger and Stunell[4]).

Ignoring the impact of localised erosion the stone has one side which would have been quite smooth. The orientation and gradient were surveyed using nine different locations on this side that were mostly untouched by erosion see Figure 2. This north-facing side is orientated so that it slopes up towards the South, its strike perpendicular to the gradient was measure to be $(92.0 \pm 2.1)°$ from geographic North and the overall gradient was $(58.3 \pm 2.9)°$. Already using only a tape measure to assess the steepness of the slope and a magnetic compass, it is possible to confirm the measurements on site. Otherwise the stone appears to be upright only with a marginal tilt of $(4 \pm 4)°$ to the West.

These surveys were followed up by a high resolution three dimension survey of the surface structure of the stone. Together with the previously surveyed orientation and a rendered model of the monolith, a realistic illumination of the standing stone could be modelled including the correct ecliptic obliquity for the proposed erection time of 2,000 BC. The overall rendered model is shown in Figure 4 for the 2 June 1pm local time with shadows showing the realistic modelling of the erosion of the stone. The insets illustrate the north-facing side looking up the slope allowing to assess the impact of erosion as well as limiting cases for the illumination during local midday: The first shadowless day for the north-facing side 2 June (19 day or 2.7 weeks from mid-summer); and mid-summer with the side illuminated by the Sun. The mid-summer case visualises how the light falling onto the north-facing side does not only pass through the runnel features created by erosion towards the right and the left of that side, but also passes over the all of the top ridge. This modelling went hand in hand with a still on-going project to gather contemporary images of the illumination of the monolith.

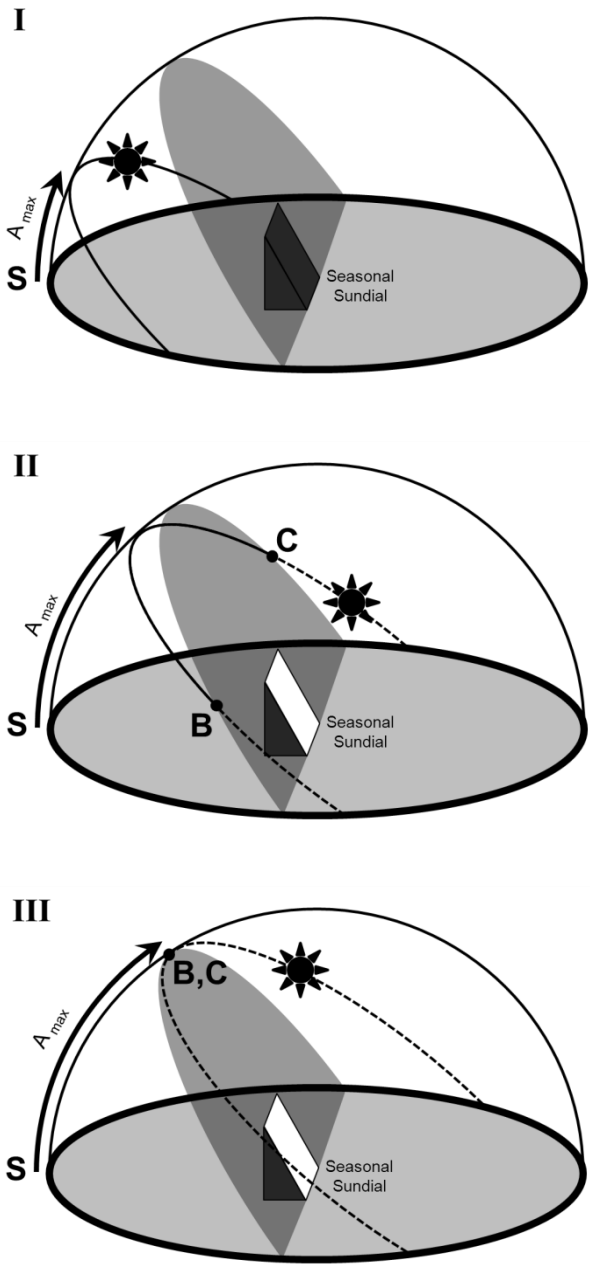

*Fig. 5. The principle of a seasonal sundial is illustrated in panels I, II, and III. The apparent path of the Sun during the winter solstice (I), an arbitrary day in the summer half of the year (II), and the summer solstice (III) is shown by the black arc with the symbol of the Sun. If the Sun is below the plane shaded in grey defined by the north-facing side of the seasonal sundial, the arc is solid. If the Sun is above the arc is dashed. The location at which the arc of the Sun intersect the plane is indicated by points B and C.*

Additionally to the work directly related to the stone, the base of the stone was also analysed. The entire site is listed and we only undertook non-invasive studies that allowed us to map the locations of any visible stones around the base of the monolith as well as measure the difference in ground elevation to a precision of 1cm. This work revealed a possible higher density of rock and boulders at the North-West base of the stone that could be linked to an increased presence of packing stones.

It is intriguing to note that the highest altitude the Sun will reach (due South) during the year for this geographic location and the time of erection of the Stone would be 60.7°. This would allow the north-facing side to be illuminated at local midday during mid-summer. Furthermore, the tilt is along the East-West axis and would not permit any light to fall onto the north-facing side during the winter-half of the year. In contrast to typical sundials it is of no interest where the shadow of the monolith or its end is located on the ground or any other structure. Treating the monolith as a gnomon or the slanted edges of the north-facing side as styles is misleading and incorrect. (Note the sloping up towards the South.) We are only analysing the illumination of the stone. In the next section a concept for a seasonal sundial will be developed allowing for seasonally relevant light and shadow casting to occur. It is in no way intended to be able to measure dates but rather illustrate the seasonally varying path of the Sun.

The principle of a seasonal sundial is modelled in Figure 5 where three panels illustrate the apparent passage of the Sun over a wedge-shaped installation with one tilted plane extended in grey. The entire illumination is illustrated with a hypothetical celestial sphere for the winter solstice (I), summer half of the year (II), and summer solstice (III). The north-facing side will be illuminated if the Sun is above the grey plane. If the Sun's path which is parallel to the celestial equator falls below the plane of the north-facing side of the stone the line is solid and if it falls above the plane it is dashed. As can be seen in Figure 5(II) the Sun can cross the plane up to two times at equal distances from the Sun's position at local midday, both marked here as points B and C. These points move closer together as the summer solstice approaches. If the plane has an obliquity regarding the horizon of less than 90-φ+ε (φ geographic latitude and ε obliquity of the ecliptic) points B and C will never merge and the north-facing side will never be illuminated during local midday. If it is exactly 90-φ+ε the points will merge at local midday and the north-facing side will always be illuminated on the day of the summer solstice as illustrated in Figure 5(III). To achieve such behaviour of the shadow, the flat side of the stone has to be orientated towards North, i.e. the plane intersects the horizon in an E-W direction. Given this behaviour, a seasonal sundial indicates the winter half of the year by having its north-facing side cast in permanent shadow. Only after the equinoxes will the north side become partly illuminated during the mornings and evenings. When the time of the summer solstice approaches the north-facing side will be illuminated during the entire day. The number of hours for which it still remains in shadow is given in Figure 6 and have been derived with the appropriate obliquity of the ecliptic. The hashed area indicates permanent illumination and the solid black area permanent shadow. The measured slope is indicated by the horizontal double line including a shaded area for the errors. The black point illustrates the result of the rendered model.

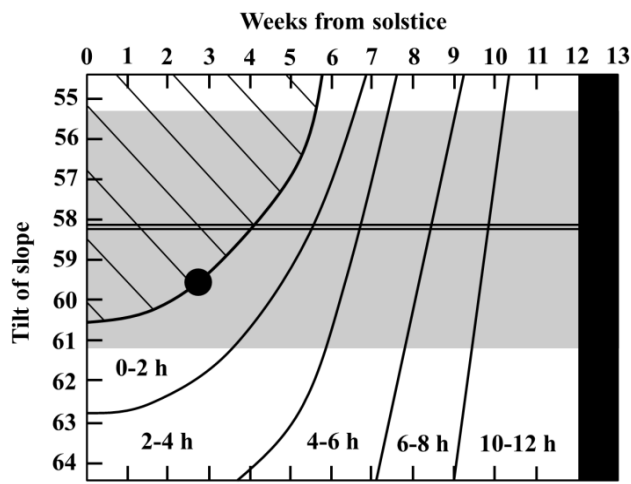

*Fig. 6. This contour map of the tilt of the plane in degrees against weeks from solstice illustrates the amount of hours the north-facing side of the stone remains in darkness around local midday. The shadow interval is binned into two hour segments, the hashed area shows times when the north side is permanently illuminated and the black area illustrates times of permanent shadow. A geographic latitude and epoch for the obliquity are chosen according to the site and erection date. The surveyed dip of the north-facing side is marked by a horizontal double line and the error range is indicated by the shaded area. The first illumination of the north side in the rendered model is marked by a black point.*

The general path of the Sun, including its varying rising and setting locations on the horizon during the year, was well known in pre-historic times. This knowledge is clearly applied in several well-known monuments across the world, including Stone Henge and other stone circles relevant for the context of the British Isles[5]. These monuments outline the typical knowledge and skills societies at the time of the erection of the Gardom's Edge monolith would have had, making its alignment achievable. However, the usage of shadows themselves to illustrate the passage of the Sun is not something so commonly encountered on the British Isles in this period. There are two sites which have been interpreted in a similar manner: Newgrange in Ireland[6] and some Clava Cairns in Scotland[7]. Both show surrounding standing stones casting a shadow onto the central burial monument during certain times that would have been of calendaric importance. But beyond that they also visualise the cyclic nature of time itself and use shadows to express eternity in a monument for the dead and living alike.

The existence of any form of calendar, megalithic or later iron-age is highly debatable[7] and of no importance for the interpretation of the proposed astronomical alignment of the monolith. It is only able to highlight the three of the four main dates during the year: equinoxes and summer solstice. Both equinoxes are also rather hypothetical in experience since their illumination effect on the stone would be only experienced during sunrise and sunset and strongly inhibited by the varying horizon and surrounding vegetation. Furthermore, determining the exact date of the summer solstices is impossible using such simple methods and was experienced in other manners. It is rather the inherent astronomical knowledge that had to be present in creating such seasonal illumination. Any other orientation could have been possible including pointing out the unmarked centre of the enclosure (azimuth (292±2)°), the rock art (azimuth (260±2)°) or a barrow known as the 'Three Men' (azimuth (282±2)°). However, these sites are all not directly observable from the standing stone. The only visible other location is the main entrance of the enclosure (azimuth (290±2)°). All of these orientations have been avoided. The finding presented here, especially the link of gradient and orientation of the slope, allows for the possibility that the standing stone at Gardom's Edge was astronomically aligned during the late Neolithic and early Bronze Age period. Therefore, this standing stone would have represented an ideal marker or social arena for seasonal gatherings for the else dispersed small communities since it incorporates seasonal shadow casting within its design. Rather than including intricate carving as indications for the seasonal importance of this stone a more natural message was incorporated into this marker. Such symbolic orientation can be compared to the alignment of mosques towards Mecca. Here the cause for seasons and life, the Sun itself, was included into the monument alignment encoding the cyclic nature of time and eternity through the delicate light and shadow play. It would have been the ideal location for ancient societies to learn more about what Francis Pryor[9] called the "lore of life" trough the landscape itself.

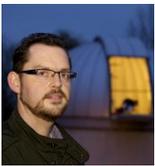

**Dr. Daniel Brown** is a professional astronomer. He is working at the Nottingham Trent University and its on-site observatory, where he supports astronomy teaching and outreach work with the general public and schools. The main focus of his outreach work is based on archaeo-astronomy and the use of the outdoor classroom. He is a founding member of the 'Horizontastronomie im Ruhrgebiet e.V.', a German private initiative promoting astronomy outreach based on an EU funded Science Park located within the Ruhr Area. He can be contacted at: daniel.brown02@ntu.ac.uk